\documentclass{article}
\usepackage{amssymb}

\usepackage{amsmath}
\usepackage{doublespace}

\input{tcilatex}

\begin{document}

\title{Alternative Descriptions in Quaternionic Quantum Mechanics\thanks{%
Work supported in part by PRIN ''Sintesi''.}}
\author{A. Blasi\thanks{%
e-mail: blasi@le.infn.it}, G. Scolarici\thanks{%
e-mail: scolarici@le.infn.it} and L. Solombrino\thanks{%
e-mail: solombrino@le.infn.it} \\
Dipartimento di Fisica dell'Universit\`{a} di Lecce \\
and INFN, Sezione di Lecce, I-73100 Lecce, Italy}
\maketitle

\begin{abstract}
We characterize the quasianti-Hermitian quaternionic operators in QQM by
means of their spectra; moreover, we state a necessary and sufficient
condition for a set of quasianti-Hermitian quaternionic operators to be
anti-Hermitian with respect to a uniquely defined positive scalar product in
a infinite dimensional (right) quaternionic Hilbert space. According to such
results we obtain two alternative descriptions of a quantum optical physical
system, in the realm of quaternionic quantum mechanics, while no alternative
can exist in complex quantum mechanics, and we discuss some differences
between them.


\end{abstract}

\section{Introduction}

Many attempts have been made in the past in order to formulate quantum
mechanics in Hilbert spaces over the skew-field $\mathbf{Q}$ of quaternions.
In the early 1960's a systematic approach began to quaternionic quantum
mechanics (\textit{QQM})\cite{fin}; at present, a clear and detailed review
of this theory, together with the foundations of quaternionic quantum field
theory, can be found in Ref.\cite{adl}.

It is worth noting that an important difference exists between complex and
quaternionic quantum mechanics about Hamiltonians operators and observables.
In both theories, observables are associated with self-adjoint (or
Hermitian) operators, whereas Hamiltonians are Hermitian in complex quantum
mechanics (\textit{CQM}), but they are anti-Hermitian in QQM, and the same
happens for the symmetry generators, like the angular momentum operators.
Moreover, in CQM any anti-Hermitian operator can be made Hermitian (and vice
versa) by multiplying by $i$. In QQM in contrast, an anti-Hermitian operator
cannot be trivially converted to a Hermitian one by multiplying by a $c$%
-number; actually in this context in order to obtain such a conversion one
needs a ''phase'' operator \cite{adl}.

Thus, if one wishes to enlarge the theoretical framework and to generalize
standard quaternionic Hamiltonians and symmetry generators (as happened in
CQM where pseudo-Hermiticity has been fruitfully introduced), in QQM\ one
rather needs to deal with \textit{pseudoanti-Hermitian} quaternionic
operators.

\textbf{Definition }\cite{sco}\textbf{.} \textit{A quaternionic linear
operator }$H$ \textit{is said to be (}$\eta $-)\textit{pseudoanti-Hermitian
if a linear invertible Hermitian operator }$\eta $ \textit{exists such that} 
\begin{equation}
\eta H\eta ^{-1}=-H^{\dagger }.
\end{equation}%
If Eq. (1) holds with a bounded positive definite $\eta $, $H$ is said 
\textit{quasianti-Hermitian}.

Of course, several $\eta $ can exist which verify Eq. (1). The properties of
pseudoanti-Hermitian Hamiltonians in QQM are analogous to the ones of
pseudo-Hermitian in CQM. In particular, a new inner product in the Hilbert
space can be associated with any bounded positive definite $\eta $ which
fulfils Eq. (1), and different $\eta $'s give rise to alternative
descriptions \cite{sim}.

In this paper, we preliminarly characterize in sec.2 the subclass of
quasianti-Hermitian quaternionic operators with discrete spectrum (in finite
dimensional vector spaces), showing that they are necessarily diagonalizable
operators with imaginary eigenvalues (and vice versa). Next, facing the
unicity problem, we derive in sec.3 a necessary and sufficient condition for
a set of quasianti-Hermitian operators to be anti-Hermitian with respect to
a uniquely defined scalar product in quaternionic Hilbert spaces. Finally,
we consider in sec.4 two alternative descriptions of a physical system in
quantum optics, which are possible only in the realm of QQM, according with
the previous result, and we discuss some differences between them.

\section{Quasianti-Hermitian quaternionic operators}

In this section, we characterize the subclass of quasianti-Hermitian
quaternionic operators by means of their spectra, in strict analogy with
similar statements in CQM \cite{ven}, \cite{scoso2}. The following
proposition, which holds in finite dimensional Hilbert spaces, provides a
necessary and sufficient condition for a quaternionic operator with discrete
spectrum to be quasianti-Hermitian.

\textbf{Proposition 1.} \textit{Let }$H$ \textit{be a quaternionic linear
operator with discrete spectrum. Then, a definite operator }$\eta $\textit{\
exists such that }$H$\textit{\ is }$\eta $-\textit{pseudoanti-Hermitian
(hence, }$\eta $-\textit{quasianti-Hermitian})\textit{\ if and only if }$H$%
\textit{\ is diagonalizable with imaginary spectrum.}

\textbf{Proof. }Let $H$ be a pseudoanti-Hermitian operator. We preliminarily
observe that, being in any case $\eta $ an invertible operator, all its
eigenvalues must be different from zero, so that either \ $\eta $ is
definite or it is indefinite. Now, let us suppose that a positive
(respectively, negative) definite operator $\eta $ exists which fulfils
condition (1); then, an $R$ exists such that $\eta =R^{\dagger }R$ \cite{zha}
(respectively, $\eta =-R^{\dagger }R$), and by Eq. (1) we obtain

\begin{equation*}
RHR^{-1}=-R^{\dagger -1}H^{\dagger }R^{\dagger }=-(RHR^{-1})^{\dagger },
\end{equation*}%
\textit{i.e.}, $RHR^{-1}$ is anti-Hermitian, hence it is diagonalizable and
it has a imaginary spectrum \cite{adl}. The same conclusion holds obviously
with regard to $H,$ since on a right quaternionic vector space the
similarity transformations preserve the properties of the spectrum, in the
sense that the real part and the moduli of the imaginary part of the
eigenvalues do not change under (quaternionic) similarity transformations.

Conversely, if $H$ is diagonalizable with imaginary spectrum, then by
proposition 2 of Ref.\cite{sco}, a positive definite operator $\eta
=SS^{\dagger }$ exists which fulfils condition (1).$\blacksquare $

We remark that the above Proposition still holds in infinite dimensional
Hilbert spaces $\mathcal{H}^{Q}$ if one assumes that the eigenvalues of $H$
have finite multiplicity and there is a basis on $\mathcal{H}^{Q}$ \ in
which $H$ is block diagonal with finite dimensional blocks (see also Ref.%
\cite{scoso2}).

As a consequence of Proposition 1, any quasianti-Hermitian operator $H$\
with discrete spectrum can be written by means of a set of biorthonormal
vectors (if we suitably fix their phases) as \cite{sco} \cite{adl}

\begin{equation*}
H=\sum_{n}\sum_{a=1}^{d_{n}}|\psi _{n},a\rangle iE_{n}\langle \phi _{n},a|,%
\text{ \ \ \ }E_{n}\geq 0,
\end{equation*}%
where $d_{n}$ denotes the degeneracy associated to the $n$th eigenvalue, $a$
is a degeneracy label and the usual relations for a biorthonormal basis hold:

\begin{equation*}
\langle \phi _{m},b|\psi _{n},a\rangle =\delta _{mn}\delta _{ba},
\end{equation*}

\begin{equation*}
\sum_{n}\sum_{a=1}^{d_{n}}|\psi _{n},a\rangle \langle \phi
_{n},a|=\sum_{n}\sum_{a=1}^{d_{n}}|\phi _{n},a\rangle \langle \psi _{n},a|=%
\mathbf{1}.
\end{equation*}

\bigskip

\section{Alternative descriptions of quantum systems}

As we already pointed out in the Introduction, different (alternative)
description of the same physical system in a quaternionic Hilbert space $%
\mathcal{H}^{Q}$ are possible whenever different $\eta $'s fulfil condition
(1). Indeed, \emph{for any bounded self-adjoint positive definite} $\eta $, 
\emph{the space }$H^{Q}$\emph{\ endowed with the scalar product }$%
\left\langle \varphi \mid \psi \right\rangle _{\eta }=\left\langle \varphi
\mid \eta \mid \psi \right\rangle $\emph{\ is a Hilbert space }$H_{\eta
}^{Q}.$

We do not report here the explicit proof of this property, which was already
stated in CQM \cite{gey}; indeed the proof easily follows from the one in
complex case since all the key steps in it\ still hold in a quaternionic
Hilbert space, as for instance the closed graph theorem \cite{semr} and the
unicity of the decomposition $\eta =S^{2}$ , with $S$ positive and
self-adjoint \cite{emch}.

Hence, an undesirable ambiguity can arise, as we will explicitly show in the
next section by means of a physical example; in order to remove that and
obtain a proper (quaternionic) quantum mechanical interpretation, we will
make resort to the concept of irreducibility of the physical operators on $%
\mathcal{H}^{Q}.$

As a preliminary step, we state the following lemma, which actually is very
similar to the quaternionic version of the corollary of the Schur Lemma (on
the irreducible quaternionic group representations of unitary operators) %
\cite{emch} and can be easily proven in the same way.

\textbf{Lemma. } \textit{Let }$\{H_{i}\}$ $(i=1,2,...,N)$ \textit{be an
irreducible set of antiself-adjoint bounded quaternionic linear operators on
the (right) quaternionic Hilbert space }$\mathcal{H}^{Q}$. \textit{Then, the
commutant of }$\{H_{i}\}$\textit{, i.e., the set of all bounded quaternionic
linear operators which commute with each }$H_{i}$\textit{\ is composed of
the operators }$T=h\mathbf{1}+a\mathbf{I}_{a}$ (\textit{where }$h,a\in 
\mathbf{R},$ \textit{and} $\mathbf{I}_{a}$ \textit{is a unitary,
anti-Hermitian operator on }$\mathcal{H}^{Q}$)\textit{.}

Thus, the following proposition provides a necessary and sufficient
condition for a set of $\eta $-quasianti-Hermitian quaternionic operators to
admit a unique positive definite operator $\eta $ which satisfy the
quasianti-Hermiticity condition.

\textbf{Proposition 2.} \textit{Let }$\{H_{i}\}$ \textit{be a set of bounded 
}$\eta $-\textit{quasianti-Hermitian operators on a right quaternionic
Hilbert space }$\mathcal{H}^{Q}$\textit{,} \textit{where }$\eta $ \textit{%
denotes a bounded positive selfadjoint operator. Then, }$\eta $\textit{\ is
uniquely determined up to a global normalization factor if and only if the
set }$\{H_{i}\}$\textit{\ is irreducible on }$\mathcal{H}^{Q}$\textit{.}

\textbf{Proof.} Firstly, we observe that, by assumption, the set of
quasianti-Hermitian observables $H_{i}$ are bounded both on $\mathcal{H}^{Q}$%
and on $\mathcal{H}_{\eta }^{Q}$, since $||H_{i}x||_{\eta }=||SH_{i}x||\leq
||SH_{i}S^{-1}||||Sx||=||SH_{i}S^{-1}||||x||_{\eta }$ (where the
decomposition $\eta =S^{2}$ , with $S$ positive, self-adjoint has been used %
\cite{emch}); furthermore, they are anti-selfadjoint on $\mathcal{H}_{\eta
}^{Q}$ because $\eta H_{i}=-H_{i}^{\dagger }\eta $ $\forall i=1,2,...,N$.
Assume now that an $\eta ^{\prime }$ exists with the same properties as $%
\eta $. Then, it follows that $[\eta ^{\prime -1}\eta ,H_{i}]=0$ $\forall
i=1,2,...,N$. Hence, by the previous lemma , $\eta =\eta ^{\prime }(h\mathbf{%
1}+a\mathbf{I}_{a})$. But imposing the Hermiticity condition on $\eta $, one
easily obtains $\eta ^{\prime }(h\mathbf{1}+a\mathbf{I}_{a})=(h\mathbf{1}-a%
\mathbf{I}_{a})\eta ^{\prime }$, which implies either $a=0$ or $\{\eta
^{\prime },\mathbf{I}_{a}\}=0$. Denoting by $|\eta ^{\prime }\rangle $ an
eigenvector of $\eta ^{\prime }$ : $\eta ^{\prime }|\eta ^{\prime }\rangle
=\alpha |\eta ^{\prime }\rangle $ (where $\alpha >0$, since $\eta ^{\prime }$
is positive) the condition $\{\eta ^{\prime },\mathbf{I}_{a}\}=0$ would
imply $\eta ^{\prime }(\mathbf{I}_{a}|\eta ^{\prime }\rangle )=-\alpha (%
\mathbf{I}_{a}|\eta ^{\prime }\rangle )$ , i.e., an eigenvector of $\eta
^{\prime }$ would exist associated with a negative eigenvalue, contradicting
thus the hypothesis on the positive definiteness of $\eta ^{\prime }$. Then, 
$a=0$ and $\eta =\eta ^{\prime }h$.

The converse is easily proven by merely paraphrasing the analogous proof in
complex Hilbert spaces \cite{gey}. $\blacksquare $

As a consequence of the above proposition, any reducible set $\{H_{i}\}$ of
quasianti-Hermitian operators admits at least two different positive
operators $\eta $ and $\eta ^{\prime }$ which fulfil the
quasianti-Hermiticity condition for any operator belonging to this set. This
allows us to construct two different Hilbert spaces $\mathcal{H}_{\eta }^{Q}$
and $\mathcal{H}_{\eta ^{\prime }}^{Q}$ , endowed with scalar products $%
\langle \varphi |\eta |\psi \rangle $ and $\langle \varphi |\eta ^{\prime
}|\psi \rangle $ respectively, such that any $H_{i}$ is anti-Hermitian on $%
\mathcal{H}_{\eta }^{Q}$ as well as on $\mathcal{H}_{\eta ^{\prime }}^{Q}$.

In particular, any reducible set of anti-Hermitian operators $\{H_{i}\}$\ on 
$\mathcal{H}^{Q}$ will appear at the same time as a set of anti-Hermitian
operators on the Hilbert space $\mathcal{H}_{\eta }^{Q}$ where $\eta $
denotes a bounded, non trivial, positive operator which commutes with any
element of $\{H_{i}\}$.

This is just the scenario of the example we will study in next section,
which exactly mimics an analogous situation in CQM, where alternative
descriptions arise in correspondence with different $\eta $'s which fulfil
the quasi-Hermiticity condition for a set $\{H_{i}\}$ \cite{sim} \cite{gey}.

\section{A physical example}

Let us consider a two level quantum optical system in the complex Hilbert
space $\mathcal{H}$ whose dynamics is described by the complex
anti-Hermitian Hamiltonian

\begin{equation}
H=2\Omega _{0}J_{1}+2\Omega _{1}J_{2}+\omega J_{3}\text{ \ \ \ }(\hbar =1),%
\text{ \ \ }\Omega _{0},\Omega _{1},\omega \in \mathbf{R},
\end{equation}%
i.e., a (real) linear combination of the anti-Hermitian operators $J_{l}$ $%
(l=1,2,3)$ , which obey the usual rules of commutation of the $SU(2)$ algebra

\begin{equation*}
\lbrack J_{l},J_{m}]=-\varepsilon _{lmn}J_{n}.
\end{equation*}%
Hamiltonian (2) can be used, for instance, in order to describe the
interaction of a chirped classical e.m. field with a two level atomic system
in a complex Hilbert space \cite{dat} . This model has been extensively
studied also to explain the Berry phase \cite{lai}.

As we already noted in the Introduction, $H$ times $i$ is of course an
observable in CQM, and it coincides with the one introduced in \cite{dat}.

By resorting to the spinorial representation of the $J$ operators

\begin{equation}
J_{1}=\frac{i}{2}\left( 
\begin{array}{cc}
0 & 1 \\ 
1 & 0%
\end{array}%
\right) ,\text{ \ \ }J_{2}=\frac{1}{2}\left( 
\begin{array}{cc}
0 & 1 \\ 
-1 & 0%
\end{array}%
\right) ,\text{ \ \ }J_{3}=\frac{i}{2}\left( 
\begin{array}{cc}
1 & 0 \\ 
0 & -1%
\end{array}%
\right)
\end{equation}%
and putting $\Omega =\Omega _{0}+i\Omega _{1}$, we can write the Hamiltonian
(2) as a $2\times 2$ anti-Hermitian (time dependent) complex matrix :

\begin{equation}
H=i\left( 
\begin{array}{cc}
\frac{\omega (t)}{2} & \Omega ^{\ast }(t) \\ 
\Omega (t) & -\frac{\omega (t)}{2}%
\end{array}%
\right) .\text{\ \ }
\end{equation}

By changing the parameters in Eq. (2) or (4), we actually obtain a set of
anti-Hermitian complex operators, which is of course irreducible in the $2$%
-dimensional (complex) Hilbert space $\mathcal{H}$, since such is the
spinorial representation of the $J_{l}$'s.

From a different point of view, we can interpret the Hamiltonian (4) as a 
\textit{anti-Hermitian quaternionic operator} in a (right) quaternionic
Hilbert space $\mathcal{H}^{Q}$, and the dynamics of our quantum system is
then described by the Schroedinger equation \cite{adl}

\begin{equation}
\frac{d}{dt}|\Psi \rangle =-H|\Psi \rangle
\end{equation}%
where $|\Psi \rangle $ belongs to $\mathcal{H}^{Q}.$ (We recall that in QQM
the eigenvalues of a anti-Hermitian Hamiltonian are imaginary quaternions,
whose moduli represent the values of the energy of the system).

Roughly speaking, $\mathcal{H}^{Q}$ can be obtained from $\mathcal{H}$ by
simply adding to each complex vector $\left| v\right\rangle \in \mathcal{H}$
a term $\left| v^{\prime }\right\rangle j$, where $j:j^{2}=-1$ is a
quaternionic unity different from $i$ ; note that $\dim \mathcal{H}^{Q}=\dim 
\mathcal{H}=2$ . Actually the various manner in which one can \textit{%
quaternionify} a complex Hilbert space are all equivalent to this one \cite%
{sh}.

Now, let us denote by $|\Psi \rangle =\left( 
\begin{array}{c}
\Psi _{\alpha ,+}+\Psi _{\beta ,+}j \\ 
\Psi _{\alpha ,-}+\Psi _{\beta ,-}j%
\end{array}%
\right) =\left( 
\begin{array}{c}
\Psi _{+} \\ 
\Psi _{-}%
\end{array}%
\right) $, $(\Psi _{\alpha ,\pm },\Psi _{\beta ,\pm }\in \mathbf{C})$ the
quaternionic state vector representing the system; the components $\Psi _{-}$
and $\Psi _{+}$ can be interpreted from a physical point of view as the
probability amplitudes for the system of being in the lowest or in the
excited state, respectively. From the Schroedinger equation one immediately
gets the time evolution of the components $\Psi _{\pm }$:%
\begin{equation}
\left\{ 
\begin{array}{c}
\Psi _{\alpha ,+}^{\prime }=\frac{i}{2}\omega (t)\Psi _{\alpha ,+}+i\Omega
^{\ast }(t)\Psi _{\alpha ,-}, \\ 
\Psi _{\alpha ,-}^{\prime }=-\frac{i}{2}\omega (t)\Psi _{\alpha ,-}+i\Omega
(t)\Psi _{\alpha ,+},%
\end{array}%
\right.
\end{equation}

\begin{equation}
\left\{ 
\begin{array}{c}
\Psi _{\beta ,+}^{\prime }=\frac{i}{2}\omega (t)\Psi _{\beta ,+}+i\Omega
^{\ast }(t)\Psi _{\beta ,-}, \\ 
\Psi _{\beta ,-}^{\prime }=-\frac{i}{2}\omega (t)\Psi _{\beta ,-}+i\Omega
(t)\Psi _{\beta ,+}.%
\end{array}%
\right.
\end{equation}%
where the prime denotes a time derivative.

Since the systems in (6) and (7) are identical, and they represent a
rotation of the vector $\Psi $ in the complex space, we can write their
solutions as a whole using the Cayley-Klein (\textit{CK}) matrix,
independently on the quaternionic or complex character of $\Psi _{\pm }:$

\begin{equation}
\left( 
\begin{array}{c}
\Psi _{+} \\ 
\Psi _{-}%
\end{array}%
\right) =\left( 
\begin{array}{cc}
F^{\ast } & G \\ 
-G^{\ast } & F%
\end{array}%
\right) \left( 
\begin{array}{c}
\Psi _{+}(0) \\ 
\Psi _{-}(0)%
\end{array}%
\right) ,\text{ \ \ }(F,G\in \mathbf{C})
\end{equation}%
where $F(t)$ and $G(t)$ are complex functions depending on $\omega $ and $%
\Omega $ in a rather involved way; furthermore $F(0)=1,$ $G(0)=0$, and $%
|F|^{2}+|G|^{2}=1$ \cite{dat}.

The CK matrix can be regarded as the matrix representation of the time
evolution operator $U$ associated with the time dependent Hamiltonian (4),
and it constitues a bi-dimensional (complex) unitary representation of the $%
SU(2)$ group.

We remark once again that the form of $U$ in (8) does not depend on the
scalar field, $\mathbf{C}$ or $\mathbf{Q}$, adopted. Now, as long as we
study the two-level system in $\mathcal{H}$, the matrix form of $U$ is
clearly irreducible, hence, by the corollary of the Schur Lemma, no
non-trivial $\eta $ exists which commutes with $U$. Recalling the discussion
at the end of previous section, we can conclude that the description of the
system in $\mathcal{H}$ is unique.

On the contrary, if we now consider $U$ as a quaternionic group
representation acting on $\mathcal{H}^{Q}$, it can be proven that this
representation is reducible into the direct sum of two equivalent
unidimensional irreducible quaternionic representations on $\mathcal{H}^{Q}$ %
\cite{fin2}, \cite{scoso}, so that $U$ admits a non-trivial commutant. By a
direct computation, the most general quaternionic Hermitian matrix $\eta $
commuting with $U$ (and $H$) is

\begin{equation}
\eta =\left( 
\begin{array}{cc}
a & jz \\ 
-jz & a%
\end{array}%
\right) ,\text{ \ }z\in \mathbf{C}.
\end{equation}
Since its matrix elements are independent of the Hamiltonian, $\eta $ is a 
\textit{secular} metric in the sense of \cite{ahmed}.

Moreover, $\eta $ is positive definite whenever $a>\left| z\right| $, as one
can prove by solving the eigenvalue problem associated with it \cite{deleo}.

We can conclude that the group representation $U$ is unitary on $\mathcal{H}%
^{Q}$

\begin{equation}
U^{\dagger }U=\mathbf{1},\text{ }
\end{equation}%
and, moreover, it is $\eta $-unitary on $\mathcal{H}_{\eta }^{Q}$ \cite{most}%
, i.e.,

\begin{equation}
U^{\dagger }\eta U=\eta .
\end{equation}

Alternatively, we can say that the Hamiltonian $H$ given in Eq. (4) is
anti-Hermitian on $\mathcal{H}^{Q}$\ as well as on the Hilbert space $%
\mathcal{H}_{\eta }^{Q}$ endowed with the scalar product $\langle \Psi |\eta
|\Phi \rangle $, since

\begin{equation}
H=\eta H\eta ^{-1}=-H^{\dagger },
\end{equation}%
where $\eta $ is given in Eq.(9).

Then, we may describe the dynamics of our system in $\mathcal{H}^{Q}$ or in $%
\mathcal{H}_{\eta }^{Q}$. Moreover, if the value $a=1$ is chosen in Eq. (9),
one obtains that for each vector $|\psi _{c}\rangle $, with \textit{complex
components} $\langle \psi _{c}|\psi _{c}\rangle $=$\langle \psi _{c}|\eta
|\psi _{c}\rangle $. The relevant physical quantities with respect to both
the alternative descriptions can now be easily computed.

Let us compute firstly the diagonal matrix elements of the angular momentum
operators and of the Hamiltonian when the system is described by the vector $%
|+\rangle =\left( 
\begin{array}{c}
1 \\ 
0%
\end{array}%
\right) $ and $|-\rangle =\left( 
\begin{array}{c}
0 \\ 
1%
\end{array}%
\right) $. (In the sequel, by an abuse of language, we will call them 
\textit{expectation values}). One easily obtains

\begin{equation*}
\langle \pm |J_{1}|\pm \rangle =0,\text{ \ \ }\langle \pm |\eta J_{1}|\pm
\rangle =\mp \frac{1}{2}kz,
\end{equation*}

\begin{equation*}
\langle \pm |J_{2}|\pm \rangle =0,\text{ \ \ }\langle \pm |\eta J_{2}|\pm
\rangle =\mp \frac{1}{2}jz,
\end{equation*}

\begin{equation*}
\langle \pm |J_{3}|\pm \rangle =\langle \pm |\eta J_{3}|\pm \rangle =\pm 
\frac{i}{2},
\end{equation*}

\begin{equation*}
\langle \pm |H|\pm \rangle =\pm \frac{i}{2}\omega ,\text{ \ \ }\langle
+|\eta H|+\rangle =\frac{i}{2}\omega -kz\Omega ,\text{ \ \ }\langle -|\eta
H|-\rangle =-\frac{i}{2}\omega -kz\Omega ^{\ast }.
\end{equation*}%
All these values are obviously imaginary quaternions. In particular the
moduli of the mean values of $H$ 
\begin{equation}
|\langle \pm |H|\pm \rangle |=\frac{|\omega |}{2},\ \ |\langle \pm |\eta
H|\pm \rangle |=\sqrt{\frac{\omega ^{2}}{4}+|z|^{2}|\Omega |^{2}},
\end{equation}%
showing then a sharp difference between the two desciptions, which however
vanishes as $\left| z\right| \longrightarrow 0.$

More generally, one can compute all the expectation values associated with
any vector $|\Psi \rangle =\left( 
\begin{array}{c}
\Psi _{+} \\ 
\Psi _{-}%
\end{array}%
\right) $, $(\Psi _{\pm }\in \mathbf{Q})$, beeing trivially $|\Psi \rangle
=|+\rangle \Psi _{+}+|-\rangle \Psi _{-}$. The only obvious warning concerns
the norm of $|\Psi \rangle $; since (as one can obtain by an easy
calculation)

\begin{equation}
\langle \Psi |\eta |\Psi \rangle =|\Psi _{+}|^{2}+|\Psi _{-}|^{2}+2\func{Re}%
\{\overline{\Psi }_{+}jz\Psi _{-}\}\neq \langle \Psi |\Psi \rangle .
\end{equation}%
(Here, $\overline{\Psi }_{+}$ denotes the quaternionic conjugate of $\Psi
_{+}$).

Finally, making resort to the form (8) of the evolution operator $U$, we can
also compute the transition probabilities in both the descriptions. Let us
for instance assume that the system is in the excited state $|+\rangle $ at $%
t=0$; the probability of finding the system in the ground state $|-\rangle $
at the time $t$ is given by

\begin{equation}
\mathcal{P}_{+\rightarrow -}(t)=|\langle -|U|+\rangle |^{2}=|G|^{2}
\end{equation}%
according to the first description, and by \cite{blasco}

\begin{equation}
\mathcal{P}_{+\rightarrow -}^{\prime }(t)=|\langle -|\eta U|+\rangle
|^{2}=|z|^{2}|F|^{2}+|G|^{2}
\end{equation}%
according to the alternative description.

We emphasize in conclusion that the possibility of an alternative
description can only occur in QQM, which then appears as a theory
intrinsecally different from CQM, and not a mere trascriptionof it.


\bigskip

\begin{thebibliography}{99}
\bibitem{fin} D. Finkelstein, J. M. Jauch, S. Sciminovich and D. Speiser, 
\textit{J. Math. Phys.} \textbf{3} (1962) 207.

\bibitem{adl} S. L. Adler, {\footnotesize \ }''\textit{Quaternionic Quantum
Mechanics and Quantum Fields}'' (Oxford University, New York, 1995).

\bibitem{sco} G. Scolarici,\textit{\ J. Phys.} A \textbf{35} (2002) 7493.

\bibitem{sim} G. Marmo, A. Simoni and F. Ventriglia, \textit{Rep. Math. Phys.%
} \textbf{51} (2003) 275.

\bibitem{ven} F. Ventriglia, \textit{Mod. Phys. Lett. }\textbf{A17} (2002)
1589.

\bibitem{scoso2} G. Scolarici and L. Solombrino, \textit{J. Math. Phys. }%
\textbf{44}\textit{\ }(2003)\textit{\ }4450.

\bibitem{zha} F. Zhang, \textit{Lin. Alg. Appl.} \textbf{251 }(1997) 21.

\bibitem{gey} F. G. Scholtz, H. B. Geyer and F. J. W. Hahne, \textit{Ann.
Phys.} \textbf{213} (1992) 74.

\bibitem{semr} P. \v{S}emrl, \textit{Commun. Math. Phys. }\textbf{242}
(2003) 579.

\bibitem{emch} G. Emch, \textit{Helv. Phys. Acta }\textbf{36} (1963) 739,
770.

\bibitem{dat} G. Dattoli, M. Richetta and A. Torre, \textit{Phys. Rev. A} 
\textbf{37 }(1988) 2007; G. Dattoli and A. Torre, \textit{J. Math. Phys.} 
\textbf{31} (1989) 236.

\bibitem{lai} Y-Z Lai, J-Q Liang, H. J. W. M\"{u}ller-Kirsten and J-G Zhou,%
\textit{\ J. Phys. A} \textbf{29 }(1996) 1773.

\bibitem{sh} C. S. Sharma, \textit{J. Math. Phys.} \textbf{29} (1988) 1069.

\bibitem{fin2} D. Finkelstein, J. M. Jauch, S. Sciminovich and D. Speiser, 
\textit{J. Math. Phys}. \textbf{4} (1963) 136.

\bibitem{scoso} G. Scolarici and L. Solombrino, \textit{J. Math. Phys. }%
\textbf{38} (1997) 1147.

\bibitem{ahmed} Z. Ahmed, \textit{J. Phys.} A \textbf{36} (2003) 9711.

\bibitem{deleo} S. De Leo, G. Scolarici and L. Solombrino,\textit{\ J. Math.
Phys.} \textbf{43} (2002) 5815.

\bibitem{most} L. Bracci, G. Morchio and F. Strocchi, \textit{Commun. Math.
Phys. }\textbf{41} (1975) 289.

\bibitem{blasco} A. Blasi, G. Scolarici and L. Solombrino, \textit{J. Phys.
A.} \textbf{37} (2004)\ 4335.
\end{thebibliography}
\end{document}